\documentclass{elsart}
\usepackage{natbib}
\begin{document}
\runauthor{Markel}
\begin{frontmatter}
\title{Comment on ``Theory of nonlinear ac responses of inhomogeneous
  two-component composite films'' [Phys. Lett. A 357, 475 (2006)].}
\author{Vadim A. Markel\thanksref{email}}
\thanks[email]{e-mail: vmarkel@mail.med.upenn.edu}

\begin{abstract}
In this comment, I point out to several mathematical mistakes in the
above-referenced letter.
\end{abstract}
\end{frontmatter}

In a recent letter~\cite{xu_06_1}, Xu, Huang and Yu (referred to as
XHY below) have derived effective nonlinear susceptibilities of graded
composite films. The theoretical approach used by XHY is based on the
perturbative theory developed previously in Ref.~\cite{yu_93_1} by Yu
(the same author as in the letter which is subject of this comment),
Hui and Stroud. However, XHY make several mathematical mistakes in
applying the formalism of Ref.~\cite{yu_93_1}, which are briefly
detailed below.

XHY have considered a two-component film whose composition is varied
in one dimension. More specifically, the volume fractions of the two
components making up the film, $p_1$ and $p_2$ ($p_1 + p_2 = 1$) were
assumed to depend on the linear coordinate $z\in [0,L]$. Nevertheless,
the computation of effective nonlinear responses, given in Eqs.~1-15
of Ref.~\cite{xu_06_1}, was carried out for fixed values of $p_1$ and
$p_2$. The dependence of effective linear and non-linear
susceptibilities on $z$ was then expressed through the dependence of
$p_1$ and $p_2$ on $z$, as is evident from Eqs.~16,17 of
Ref.~\cite{xu_06_1}. Thus, the gradation of the films was assumed to
be slow enough, so that the effective susceptibilities could be
physically defined as functions of $z$.  This assumption can be
reasonable and is not subject of this comment. However, the method
used for the derivation of the effective constants {\em at fixed
  values of $p_1$ and $p_2$} outlined in Eqs.~1-15 of
Ref.~\cite{xu_06_1} is erroneous.

The derivation mentioned above is based on Eq.~14 of
Ref.~\cite{yu_93_1}. This equation gives a formula for computing the
effective linear dielectric constant of a composite, $\epsilon_e$,
and, written in terms of the volume fractions $p_1$ and $p_2$ and the
linear dielectric constants $\epsilon_1$ and $\epsilon_2$ of the
components, has the form

\begin{equation}
\label{main}
\epsilon_e = p_1 \epsilon_1 \frac{\langle E_1^2
  \rangle}{E_0^2} + p_2 \epsilon_2 \frac{\langle E_2^2
  \rangle}{E_0^2} \ .
\end{equation}

\noindent
In this formula, $\langle E_1^2 \rangle$ and $\langle E_2^2 \rangle$
are the averages of the square of the (real-valued) linear electric
field, computed inside the first and the second constituent of the
film, respectively, and $E_0$ is the external electric field. Because
$E_1$ and $E_2$ are {\em linear} fields, the ratios $\langle E_{1,2}^2
\rangle/E_0^2$ are independent of $E_0$. Thus, Eq.~\ref{main} gives
the effective linear dielectric constant of the composite in terms of
the linear dielectric constants of its constituents as a weighted
average.  Incidentally, this definition of $\epsilon_e$ was obtained
by equating the total electrostatic energy of a homogeneous sample
occupying some volume $V$ and characterized by the linear dielectric
constant $\epsilon_e$ and that of a composite occupying the same
volume and characterized by the constants $p_{1,2}$ and
$\epsilon_{1,2}$. Of course, the ratios $\langle E_{1,2}^2
\rangle/E_0^2$ must be computed by solving the electrostatic boundary
value problem for each specific geometry of the composite.

In what follows, I detail three mistakes XHY have made in applying
Eq.~\ref{main} to the the problem of computing the nonlinear responses
of composite films.

\paragraph*{The First Mistake} 
At the onset, XHY use Eq.~\ref{main} to compute the averages $\langle
E_1^2 \rangle$ and $\langle E_2^2 \rangle$, assuming that $\epsilon_e$
is given by some known function, $\epsilon_e = F(\epsilon_1,
\epsilon_2, p_1, p_2)$. Namely, they write (Eqs.~2,3,7 of
Ref.~\cite{xu_06_1}):

\begin{equation}
\label{averages}
\langle E_1^2 \rangle = \frac{1}{p_1}\frac{\partial F(\epsilon_1,
  \epsilon_2, p_1, p_2)}{\partial \epsilon_1} E_0^2 
\end{equation}

\noindent
and analogously for $\langle E_2^2 \rangle$. In writing this equation,
XHY have ignored the fact that the averages $\langle E_1^2 \rangle$
and $\langle E_2^2 \rangle$ are themselves functions of both
$\epsilon_1$ and $\epsilon_2$. Differentiation of Eq.~\ref{main} with
respect to $\epsilon_1$ (taking into account $\epsilon_e=F$) would
yield

\begin{equation}
\label{derivative}
p_1 \left( \langle E_1^2 \rangle + \epsilon_1 \frac{\partial \langle
    E_1^2 \rangle}{\partial \epsilon_1} \right) + p_2 \epsilon_2
\frac{\partial \langle E_2^2 \rangle}{\partial \epsilon_1} =
\frac{\partial F(\epsilon_1, \epsilon_2, p_1, p_2)}{\partial \epsilon_1} E_0^2 \ .
\end{equation}

\noindent
Eq.~\ref{averages} is derivable from Eq.~\ref{derivative} only if
$\partial \langle E_1^2 \rangle/\partial\epsilon_1 = \partial \langle
E_2^2 \rangle/\partial\epsilon_1 = 0$, which, obviously, is not the
case.

\paragraph*{The Second Mistake}
Application of Eq.~\ref{averages} requires the knowledge of the
function $F(\epsilon_1, \epsilon_2, p_1, p_2)$. To this end, XHY
define $F$ for the graded film as a whole (not locally) by writing
$F=L[\int_0^L \epsilon_{\rm MG}^{-1}(z) dz]^{-1}$ (Eq.~21 of
Ref.~\cite{xu_06_1}), where $\epsilon_{\rm MG}$ is the Maxwell-Garnett
effective dielectric constant that can be found analytically from
Eqs.~16,17 of Ref.~\cite{xu_06_1}. The first step in this procedure
has not been justified in Ref.~\cite{xu_06_1} and appears to be
arbitrary. More importantly, the second step requires that the
Maxwell-Garnett formula give the same result for the dielectric
constant as Eq.~\ref{main} with properly computed field averages.  But
the two equations are, generally, not equivalent. In
Ref.~\cite{yu_93_1}, a Maxwell-Garnett-type formula was derived from
Eq.~\ref{main} for the case of spherical inclusions of volume fraction
$p_1$ in the limit $p_1 \rightarrow 0$. But for mixing ratios close to
$0.5$, and for high-quality metal inclusions in a dielectric host with
the electromagnetic frequency being close to the Frohlich resonance of
a single inclusion (all of which is the case in numerical examples
shown in Ref.~\cite{xu_06_1}), the Maxwell-Garnett formula is known to
be very inaccurate~\cite{stockman_99_1}. In fact, it has been
demonstrated in Ref.~\cite{stockman_99_1} that the Maxwell-Garnett
theory provides a reasonable approximation only for $p_1<10^{-3}$.

Physically, computing the averages $\langle E_1^2 \rangle$ and
$\langle E_2^2 \rangle$ by means of Eq.~\ref{averages} (which is, also
incorrect due to the First Mistake), where the analytical form of $F$
is derived from the Maxwell-Garnett formula ignores the well-known
phenomenon of strong fluctuations of electric field in resonant
composites~\cite{stockman_00_2,sarychev_00_1}.

\paragraph*{The Third Mistake}
XHY make the most serious mistake when they state that the {\em
  nonlinear} susceptibilities can be obtained from Eq.~\ref{main} by
viewing the constants $\epsilon_{1,2}$ as intensity-dependent, i.e.,
by making the substitutions $\epsilon_{1,2} \rightarrow \epsilon_{1,2}
+ \chi_{1,2} \langle E_{1,2}^2 \rangle$ in the arguments of the
function $F$, and by expanding $F$ with respect to the small
parameters $\chi_{1,2}$. This procedure is mathematically expressed in
Eq.~6 of Ref.~\cite{xu_06_1} and is clearly erroneous.  By definition,
$\epsilon_{1,2}$ in Eq.~\ref{main} are the {\em linear} dielectric
constants of the composite constituents, as well as $\epsilon_e$ is,
by definition, the linear effective dielectric constant of the
composite. Computation of higher-order effective susceptibilities
would require computing higher moments of the field.  For example, the
formula derived in Ref.~\cite{yu_93_1} for the effective third-order
susceptibility is

\begin{equation}
\label{third_order}
\chi_e = p_1 \chi_1 \frac{\langle E_1^4
  \rangle}{E_0^4} + p_2 \chi_2 \frac{\langle E_2^4
  \rangle}{E_0^4} \ .
\end{equation}

\noindent 
Thus, Eq.~6 of Ref.~\cite{xu_06_1} is based on an incorrect
interpretation of the theoretical results of Ref.~\cite{yu_93_1}

Finally, the conclusion of Ref.~\cite{xu_06_1} that ``the harmonics
[generated in a graded composite film] are significantly dependent on
the gradation profiles as well as the width of the composite film'' is
technically (and trivially) correct. However, equations derived in
Ref.~\cite{xu_06_1} can not be used to ``monitor the gradation profile
as well as the width of the composite graded film by measuring the
nonlinear ac responses of the film subjected to an ac electric field''
due to the errors described in this comment.

\bibliographystyle{elsart-num}
\bibliography{abbrev,master}

\end{document}